\documentclass[12pt]{iopart}
\usepackage{latexsym}

\def\ran{{\rm ran}}
\def\span{{\rm span}}
\def\supp{{\rm supp}}
\def\ket#1{\mid~\!\!\!{#1}~\!\!\rangle}
\def\bra#1{\langle~\!\!{#1}~\!\!\!\mid}
\def\cH{{\cal H}}
\def\cS{{\cal S}}

\begin{document}\jl{1}

\title[\bf Different State Assignments] {\bf
On Compatibility and Improvement of
Different Quantum State Assignments}
\author{F Herbut\footnote[1]{E-mail:
fedorh@infosky.net}}
\address{Serbian Academy of
Sciences and Arts, Knez Mihajlova 35,
11000 Belgrade, Serbia and Montenegro}

\date{\today}

\begin{abstract}
When Alice and Bob have different quantum
knowledges or state assignments (density
operators) $\rho_A$ and $\rho_B$
respectively for one and the same
specific individual system, then the
problems of compatibility and pooling
arise. The so-called first
Brun-Finkelstein-Mermin (BFM) condition
for compatibility is reobtained in terms
of possessed or sharp (i. e., probability
one) properties. The second BFM condition
is shown to be generally invalid in
infinite dimensional state space. An
argument leading to a procedure of
improvement of $\rho_A$ on account of
$\rho_B$ and {\it vice versa} is
presented.
\end{abstract}

\maketitle

\normalsize

\rm \section{Introduction}

To my knowledge, the problem of
compatible state assignments originated
with Peierls \cite{Peierls}. His
necessary conditions were seriously
criticized be Fuchs and Mermin
\cite{Mermin}. Then Brun, Finkelstein and
Mermin (BFM) \cite{BFM}, \cite{Brun}
derived two necessary conditions for {\it
compatibility} of different state
assignments $\rho_A$ and $\rho_B$, i. e.,
for two density operators, describing
(being the quantum knowledge about) one
and the same system. The two conditions
were found to be equivalent in finite
dimensional state space (to which they
confined their discussion).

{\it The first BFM condition} reads that
the intersection of the supports is at
least one dimensional:
$$dim\{\supp(\rho_A)\cap \supp(\rho_B)\}
\geq 1.\eqno{(1)}$$ (Support of a density
operator is the subspace spanned by the
eigensubspaces corresponding to positive
eigenvalues.)

{\it The second BFM condition} states
that there exist pure-state expansions
$$\rho_A= p_A\ket{\phi }\bra{\phi }+
\sum_{i\geq
1}p_{Ai}\ket{\phi_{Ai}}\bra{\phi_{Ai}},$$
$$\rho_B=p_B\ket{\phi }\bra{\phi }+
\sum_{i\geq 1}p_{Bi}\ket{\phi_{Bi}}
\bra{\phi_{Bi}}\eqno{(2)}$$ (with all
weights non-negative and $p_A$ and $p_B$
both positive) having a common pure state
$\ket{\phi }$.

In reference \cite{CFS} two approaches to
compatibility are discussed. One is shown
to be equivalent to the BFM condition.
The other leads to a hierarchy of
measurement-based compatibility criteria,
all inequivalent with the BFM condition.

In reference \cite{PB-K} a more general
approach based on a measure of the
compatibility between two state
assignments is expounded. This measure is
then applied to a procedure of pooling
information.

In reference \cite{Jacobs} classical and
quantum pooling of informations is
discussed. The author claims that in the
quantum case Alice and Bob must also
possess information about how their
respective states of knowledge were
obtained.

Most likely there are some more or less
important contributions that I have
unintentionally omitted in this very
short review because I am not aware of
them.

In \cite{Brun} a thorough list of further
questions is given. Clearly, the problem
of compatibility and of pooling of
information from different state
assignments (for one and the same
individual system) is not quite near to
its complete solution.

Henceforth the state space is allowed to
be finite or countably infinite
dimensional.

This article is organized as follows: In
section 2 the first BFM necessary
condition is reobtained in a
mathematically slightly but physically
considerably different way than in the
BFM article \cite{BFM}. (The approach of
section 2 is required for section 4.) In
section 3 it is shown that in infinite
dimensional state space the two BFM
conditions are not equivalent, and that
only the first one is generally valid. In
section 4 a possible improvement of
$\rho_A$ on account of $\rho_B$ and {\it
vice versa} is expounded, and a simple
way of pooling information from the two
state assignments is given. Finally, the
results of the article are summed up in a
conclusion.

Needless to say that extension of the
arguments of this article from two to any
finite number of state assignments is
straightforward.

\section{Derivation of the First BFM
Condition}

When one is dealing with statistical
knowledge, as it is the case with density
operators, then one has in mind a {\it
random} element of the ensemble. As far
as a specific individual system from the
ensemble is concerned, the statistical
notions like the average hardly make
sense.

If $P$ is a projector (physically:
property or event), then $TrP\rho$ is the
{\it probability} of possession of the
property by (or of occurrence of the
event on) a random system from the
ensemble described by $\rho$. In the
special case when $\Tr (P\rho )=1$, i.
e., when one has a {\it sharp} or
possessed {\it property} (a certain
event), it is easy to show that for any
state decomposition $\rho
=\sum_kw_k\rho_k$ ($\forall k:\enskip
w_k>0, \rho_k>0, \Tr \rho_k=1;
\sum_kw_k=1$), all substates $\rho_k$
"inherit" the sharp property: $\forall
k:\enskip \Tr (P\rho_k)=1$. Analogous
statements hold true for finite
laboratory ensembles and subensembles
that represent empirically the density
operators. Hence, in terms of sharp
properties, one can speak of {\it
individual-system knowledge}, because it
applies not only to a random, but also to
a specific system in the ensemble.

If $P$ and $P'$ are two projectors
(commuting or not), one must clarify in
what case they can both be simultaneously
sharp properties of one and the same
system.

It is shown in the Appendix that $P$ is a
{\it sharp property} in the state $\rho$
{\it if and only if}:
$$ P\geq Q\quad \mbox{meaning} \quad
PQ=Q, \eqno{(3)}$$ where $Q$ projects
onto the support of $\rho$. (The first
relation expresses implication in the
lattice of projectors and the second is
its algebraic equivalent).

Condition (3) makes it obvious that two
properties $P$ and $P'$ can both be
simultaneously sharp properties of one
and the same system {\it if and only if
their greatest lower bound} $P_{glb}$
{\it is nonzero}, because then and only
then do they have a common nonzero lower
bound $Q$, which can be the support
projector of a density operator. If
$[P,P']=0$, then $P_{glb}=PP'$.

The following claim gives {\it physical
meaning to the greatest lower bound}
$P_{glb}$.

{\bf Lemma 1.} {\it Properties $P$ and
$P'$ are {\bf sharp properties} in a
state $\rho$ {\bf if and only if so is
their greatest lower bound}} $P_{glb}$.

{\bf Proof} follows immediately from the
necessary and sufficient condition (3),
because the two projectors have $Q$, the
range projector of $\rho$, as their
common lower bound if and only if
$P_{glb}\geq Q$.\hfill $\Box$

Thus, $P$ and $P'$ as sharp properties
can be replaced by the single sharp
property $P_{glb}$.

Returning to two state assignments
$\rho_A$ and $\rho_B$ concerning one and
the same system, the corresponding
support projectors $Q_A$ and $Q_B$ are
both sharp properties of the system at
issue as seen from (3). Then so is their
greatest lower bound $Q_{glb}$, and it
must not be zero, because zero cannot be
a sharp property (cf (3)). $Q_{glb}$
projects onto $\supp(\rho_A)\cap
\supp(\rho_B)$. Therefore, this subspace
must not be zero either. This is the
first BFM condition. It is obviously
valid both in finite and in infinite
dimensional state spaces.

\section{The Second BFM Condition in an
Infinite Dimensional State Space}

An important result of Hadjisavvas
\cite{Hadji} establishes the following
claim.

{\bf Lemma 2.} {\it A pure state
$\ket{\phi }$ can appear in a state
decomposition of a given density operator
$\rho$ (cf (2)) in a state space of
finite or infinite dimension {\bf if and
only if}}
$$\ket{\phi }\in
\ran(\rho^{1/2}),\eqno{(4)}$$ where
$ran(\dots )$ denotes the range.

If $\supp(\rho )$ is finite dimensional,
then $\ran(\rho^{1/2})=\ran(\rho
)=\supp(\rho)$. But if $\supp(\rho )$ is
infinite dimensional, then $$\ran(\rho
)\subset \ran(\rho^{1/2})\subset
\supp(\rho )\eqno{(5)}$$ (proper
subsets).

Let us take a simple example in which
$\rho_A$ has an infinite-dimensional
range and
$$\ket{\phi }\in \Big(\supp(\rho_A)\ominus
\ran(\rho_A^{1/2})\Big),\quad \bra{\phi }
\ket{\phi }=1.\eqno{(6a)}$$ Let, further,
$$\ket{\phi }\in
\supp(\rho_B).\eqno{(6b)}$$ Let, finally,
$$\Big(\supp(\rho_B)\ominus \span(\ket{\phi
})\Big)\perp \Big(\supp(\rho_A)\ominus
\span(\ket{\phi })\Big).\eqno{(6c)}$$
Then the first BFM condition is
satisfied, but $\ket{\phi }$, the only
common state vector (up to a phase
factor) in the supports, cannot appear in
a decomposition like the first one in (2)
on account of Lemma 2. Therefore, the two
BFM conditions are not equivalent if the
support of at least one of the state
assignments has an infinite dimensional
support.

In view of Lemma 2, the second BFM
condition is equivalent to
$$dim\{\ran(\rho_A^{1/2})\cap \ran
(\rho_B^{1/2})\}\geq 1.\eqno{(7)}$$
irrespectively of the dimensions of the
supports. If both ranges are finite
dimensional, (7) equals (1). If at least
one of the ranges is infinite
dimensional, the linear manifolds in (7)
are proper subsets of the topologically
closed subspaces appearing in (1). Then,
(7) is stronger than (1), i. e., the
former implies the latter, and I am not
aware of any argument so far that would
prove the validity of (7) as a necessary
condition for compatibility of the two
state assignments.

\section{How Two Compatible State Assignments
Can Improve Each Other}

We assume that (1) is valid, i. e., that
the two state assignments are compatible.
It may happen that $Q_{glb}\equiv
glb(Q_A,Q_B)<Q_A$, i. e., that $Q_{glb}$
is not a sharp property of the system at
issue according to $\rho_A$, though it is
known to be if also the information from
$\rho_B$ is taken into account. In this
case $\rho_A$ contains desinformation as
far as the individual system under
consideration is concerned, and one may
like to dispense with it. We lean on the
following mathematical facts in finding a
way to do so.

{\bf Lemma 3.} {\it Let  $P$ and $\rho$
be a projector and a density operator
such that $p\equiv \Tr (\rho P)>0$. Then
$\rho_L\equiv P\rho P/p$ is {\bf closest}
to $\rho$ in the sense of Hilbert-Schmidt
distance in comparison with all density
operators for which $P$ is a sharp
property.}

The capital $L$ in the index is due to my
liking to call $\rho_L$ a L\"{u}ders
state. The reader is familiar with it in
the context of change of state in ideal
measurement. To my knowledge, it was
introduced by L\"{u}ders \cite{Lud} (and
not by von Neumann \cite{vN} as many seem
to think, cf also \cite{FH2}). Its above
claimed meaning concerning distance was
established (in case of so-called
non-selective measurement, when all the
results of the measurement are taken into
account) in previous work \cite{FH1}.

{\bf Proof} of Lemma 3 follows
immediately if one takes into account the
following facts: (i) All density
operators are Hilbert-Schmidt ones, i.
e., they are elements of the Hilbert
space $\cH_{HS}$ of Hilbert-Schmidt
operators. These are linear operators $A$
such that $\Tr (A^{\dagger}A)<\infty$.
(If the state space is finite
dimensional, then all linear operators
are Hilbert-Schmidt ones.) (ii) The
superoperator $P\dots P$ is a projector
in $\cH_{HS}$. (iii) The projection of a
given vector into a given subspace of a
(complex or real) unitary space has the
smallest distance from the given vector
in comparison with all vectors from the
subspace.\hfill $\Box$

The established claim of being "closest"
may carry a mathematical elegance, but
its physical meaning may be not so
transparent. Therefore, we approach the
L\"{u}ders state from another angle.

If $\rho_A$ is to be changed into another
density operator describing a state in
which a given property $P$ will be
possessed, the statistical predictions
will change in general. Still, there can
be a set o predictions that should not
change: those properties $P'$ that imply
$P$, so that when they become possessed,
$P$ remains possessed.

{\bf Lemma 4.} {\it Let again $P,\rho$ be
given with $p\equiv \Tr (\rho P)>0$. Let,
further, $\cS\equiv \{P':P\geq P'\}$ be
the set of all projectors implying $P$.
Then $$\Tr (\rho P')=p[\Tr (\rho'P')]
\eqno{(8)}$$ for all $P'\in \cS$ {\bf if
and only if} $\rho' =\rho_L$, where
$\rho_L$ is the L\"{u}ders state (cf
Lemma 3).}

To my knowledge, a lemma related to the
claim of Lemma 4 was first proved by Bell
and Nauenberg in \cite{Bell}. We'll
resort to their argument in the proof
that follows.

{\bf Proof} of Lemma 4. Let $\cS'$ be the
subset of $\cS$ containing all its
projectors onto one-dimensional
subspaces. They can be written as
$\ket{\psi }\bra{\psi }$. Then one can
argue as follows
$$\forall \ket{\psi }\bra{\psi }\in \cS':
\quad \Tr [\rho (\ket{\psi }\bra{\psi
})]=p\{\Tr [\rho'(\ket{\psi }\bra{\psi
})]\}\quad \Leftrightarrow \bra{\psi
}\rho \ket{\psi }=p\bra{\psi
}\rho'\ket{\psi }.$$ On the other hand,
$$P\ket{\psi }=P\bra{\psi }\ket{\psi
}\ket{\psi }=\bra{\psi }\ket{\psi
}\ket{\psi }=\ket{\psi }.$$ Equivalently,
$$\ket{\psi }\in \supp(P).\eqno{(9)}$$
Obviously, (9) is not only necessary, but
also sufficient for $\ket{\psi }\bra{\psi
}\in \cS'$.

Hence, the above chain of equivalences
can be continued as follows.
$$\Leftrightarrow \bra{\psi } (P\rho
P)\ket{\psi }=p\bra{\psi }\rho' \ket{\psi
}.$$ Since $P\rho P$ is zero in the
orthocomplement of $\supp(P)$, and
$\ket{\psi }$ is an arbitrary state
vector in this subspace, we finally have
$$\rho'=P\rho P/p$$ as claimed.\hfill
$\Box$

One might still object that Lemma 4 tells
about statistical predictions (that
should not change). We have a fixed
individual system from the ensemble
described by $\rho_A$ in mind. Statistics
may not be quite applicable. Let us
return to the sharp properties. They do
have individual-system meaning.

{\bf Lemma 5.} {\it Let $\bar \cS$ be the
set of all sharp properties $P'$ in a
given state $\rho$ and let $P$ be a
statistically possible but not
necessarily sharp property of the random
system in the state $\rho$, i. e., let
$p\equiv \Tr (\rho P)>0$. Let, finally,
$\bar \cS'$ be the subset of $\bar \cS$
containing all $P'$ {\bf compatible} with
$P$ as obseravables, i. e., for which
$[P',P]=0$. Then both $P$ and each $P'$
from $\bar \cS'$ are sharp properties in
a state $\rho'$ {\bf if and only if}
$\rho'=P\rho P/p.$}

{\bf Proof.} The claim of Lemma 5 is a
special case of a wider claim proved in
\cite{FH2} as Theorem 1 there. (The
context was ideal measurement. But this
was not relevant for the somewhat
intricate proof given there.)

Returning to the two state assignments
$\rho_A$ and $\rho_B$ and to the
desinformation in the former, in view of
Lemmata 3, 4, and 5, the desinformation
can be dispensed with or $\rho_A$ can be
improved if it is replaced by
$$\bar \rho_A\equiv Q_{glb}\rho_A Q_{glb}/
p_{glb}^A, \eqno{(10)}$$ where $Q_{glb}$
is the greatest lower bound (in the
lattice of projectors) of $Q_A$ and
$Q_B$, the support projectors of $\rho_A$
and $\rho_B$ respectively, and $p_{glb}^A
\equiv \Tr (\rho_AQ_{glb})$. This
probability is necessarily positive as
proved in what follows.

{\bf Lemma 6.} {\it If $\rho$ is a
density operator with $Q$ as its support
projector, and $P$ is another nonzero
projector implying $Q$, then the
probability $\Tr (\rho P)$ is positive.}

{\bf Proof.} Let us start {\it ab
contrario} assuming that $\Tr (\rho
P)=0$. We show that $P$ is then
necessarily a subsprojector of or,
equivalently, that it implies the null
projector $(1-Q)$ of $\rho$, in
contradiction to the assumptions in Lemma
6. To prove this, we write down a
spectral form $\rho
=\sum_ir_i\ket{i}\bra{i}$ of $\rho$ in
terms of its positive eigenvalues and the
corresponding eigenvectors, and
analogously for $P$:
$P=\sum_k\ket{k}\bra{k}$. Then $$0=\Tr
(\rho
P)=\sum_i\sum_kr_i|\bra{i}\ket{k}|^2.$$
Since all terms are non-negative, all
$\ket{k}$ must be orthogonal to all
$\ket{i}$.\hfill $\Box$

Let us take stock of what has been
achieved.

\section{How Much Improvement Has Been
Achieved?}

Naturally, the symmetric expression to
(10), i. e., $\bar \rho_B=Q_{glb}\rho_B
Q_{glb}/p^B_{glb}$ with $p^B_{glb}\equiv
\Tr (\rho_BQ_{glb})$, is the improvement
of $\rho_B$. It is desirable to clarify
if there are {\it distinct sharp
properties} in $\bar \rho_A$ and $\bar
\rho_B$. Utilizing criterion (3), we
take resort to the corresponding support
projectors $\bar Q_A$ and $\bar Q_B$.

{\bf Theorem 1.} {\it The improved states
$\bar \rho_A$ and  $\bar \rho_B$ have
{\it one and the same support projector},
i. e., $\bar Q_A=\bar Q_B=Q_{glb}$.}

{\bf Proof.} It is obvious from (10) that
$Q_{glb}\bar \rho_A=\bar \rho_A$. Taking
the trace, we see that $Q_{glb}$ is a
sharp property in the state $\bar
\rho_A$. From (3) {\it mutatis mutandis}
it follows that $Q_{glb}-\bar Q_A$ is a
projector. Evaluating $\Tr [(Q_{glb}-\bar
Q_A)\bar \rho_A]$, we obtain zero. On the
other hand, due to $(Q_{glb}-\bar Q_A)
\leq Q_{glb}$, or equivalently
$(Q_{glb}-\bar Q_A)Q_{glb}=(Q_{glb}-\bar
Q_A)$, due to (10), and commutation under
the trace, we, further, obtain
$$0=\Tr [(Q_{glb}-\bar Q_A)\bar \rho_A]= \Tr
[(Q_{glb}-\bar Q_A)\rho_A].\eqno{(11)}$$

Taking into account that $(Q_{glb}-\bar
Q_A)\leq Q_{glb}$ and the latter is a
subprojector of the support projector of
$\rho_A$, hence also $(Q_{glb}-\bar Q_A)$
is a subprojector of the same, we see
that relation (11) and Lemma 6 imply
$Q_{glb}-\bar Q_A=0$. \hfill $\Box$

Thus, the two improved states have the
same set of sharp properties. The method
of sharp properties applied so far cannot
lead us any further. This all generalizes
trivially to the case of several state
assignments.

If one of the improved state assignments
turns out to be pure, then all are, and
all are the same pure state. With less
luck in the pooling performed so far, one
could end up with mixed improved states,
but still all equal ones. This would also
end the necessity for further pooling.
But in general the improved states can be
mixed and distinct. Then further pooling
is required.

A simple way of pooling information from
two state assignments goes as follows.
Let $0<w<1$. The result of pooling is
obtained by averaging
$$\rho \equiv
wQ_{glb}\rho_AQ_{glb}/p_{glb}^A+
(1-w)Q_{glb}\rho_BQ_{glb}/p_{glb}^B.
\eqno{(12)}$$ If the "quantum knowledges"
of both Alice and Bob are believed
equally trustworthy, then $w=1/2$ seems
in order.

 A more sophisticated way of pooling is
required in some cases (see e. g.
\cite{PB-K}, \cite{Jacobs}).

{\it CONCLUSION.} The basic method of
this article is that of {\it sharp
properties}. These are interpreted as
being possessed by the individual quantum
system about which Alice and Bob have
quantum knowledges or state assignments.
Making use of this method, the first BFM
necessary condition (see the
Introduction) was rederived. It was shown
that in case of infinite-dimensional
ranges of the density operators, the
first BFM condition is valid; the second
need not be. Finally, the method was
utilized to improve Alice's and Bob's
state assignments on account of
information from each other. Thus, they
end up with possibly different density
operators, but they have one and the same
set of sharp or possessed properties.\\

{\bf Appendix}\\

We prove now that for a projector $P$ and
a statistical operator $\rho$ with the
range projector $Q$ one has $P$ as a {\it
sharp property}, i. e., $TrP\rho =1$,
{\it if and only if} $P\geq Q$.

    {\it Sufficiency.} The assumed relation
$P\geq Q$ means $PQ=Q$. Further, $Q\rho
=\rho$. Hence,
$$1=\Tr \rho =\Tr (Q\rho )=\Tr (PQ\rho )=\Tr (P\rho ).$$

    {\it Necessity.} Let
$$ \rho =\sum_ir_i\ket{i}\bra{i},\quad
\forall i:\enskip r_i>0,$$ be a spectral
form of $\rho$. By assumption, now one
has
$$1=\sum_ir_i\Tr (P\ket{i}\bra{i})
=\sum_ir_i\bra{i}P\ket{i}.$$ Subtracting
this from $1=\sum_ir_i$, one obtains
$$0=\sum_ir_i(1-\bra{i}P\ket{i}). $$
Since always $0\leq \bra{i}P\ket{i}\leq
1$, we have
$$\forall i:\quad \bra{i}P\ket{i}=1,$$
or equivalently,
$$\forall i:\quad \bra{i}P^{\perp}\ket{i}
=0\enskip \Leftrightarrow
||P^{\perp}\ket{i}||^2=0\enskip
\Leftrightarrow P^{\perp}\ket{i}=0$$
$$\Leftrightarrow
P\ket{i}=\ket{i}.$$ Since, $\sum_i\ket{i}
\bra{i}=Q$, we,
finally, have $PQ=Q$ as claimed.\hfill $\Box$\\

\end{document}